\def \half { {1 \over 2}} 
 \def \noi {\noindent}
\def \st {_{\scriptscriptstyle T} }
\def \sl  {_ {\scriptscriptstyle L}  } 
\def \zero {_ {(0)} } 
\def \zer  {^  {(0)} } 
\def  \noi  {\noindent }
\def  \per  { _ \perp }

\def \dzy {5}       \def \rep {6}
          \def \coupl {7}  
\def \sazjmp {8}    
        \def  \bij89 {9}        \def  \dv95 {10}      \def \purmagn {11} 
\def \rst {12}               \def \rigo {13}
     \def \L.L  {14}     \def \anal {15}           \def    \sazdj  {16}      
      \def  \klambda  {17}       \def  \kel {18}
\def \wein {19}        
\def  \pilk {20}
\def \dixm {21}       \def \algo {22}             \def \brok {23}
$$ \   $$
\centerline   {\bf   Relativistic quantum mechanics}
\centerline   {\bf   of a neutral two-body system}
\centerline   {\bf in a constant magnetic field}  
\bigskip
\centerline {\bf Philippe Droz-Vincent}
\bigskip
\centerline {Laboratoire de Gravitation et Cosmologie Relativistes}
\centerline {C.N.R.S. URA 769, Universit\'e Pierre et Marie Curie}
\centerline  {Tour 22-12 , boite courrier 142}
\centerline {4 place Jussieu 75252 Paris Cedex 05, France}
\bigskip
\noi
{\it  A 
(globally) neutral two-body system is supposed to obey a pair of coupled 
Klein-Gordon equations in a constant homogeneous magnetic field.
Considering eigenstates of the pseudomomentum four-vector,
 we reduce these equations to a three-dimensional eigenvalue problem.

The frame adapted to pseudomomentum has in general a nonvanishing velocity 
with respect to the frames where the field is purely magnetic.
This velocity plays a crucial role in the occurance of motional terms;
these terms are taken into account within a  manifestly covariant framework.
 
Perturbation theory is available when the mutual interaction doesnot depend 
on the total energy; a weak-field-slow-motion 
approximation is more specially tractable}.                               


\vfill \eject

      {\bf 1. INTRODUCTION. }
 
From a relativistic point of view, the constant magnetic field has this 
peculiarity that it does not correspond to a unique laboratory frame.
When  a constant homogeneous electromagnetic field
 is seen as purely magnetic in some frame (conventionally referred to as 
laboratory frame),  such a frame cannot be unique [1] .
 The directions eligible for 
the time axis of a lab frame span a 2-dimensional hyperbolic space $(E \sl ) $ 
(longitudinal space). 
Energy, defined as time component of the momentum, is conserved 
but it is  affected by this  ambiguity about  the lab frame.
Therefore a manifestly covariant treatment is specially relevant.

 Another  important feature of the constant magnetic field is the conservation 
of pseudomomentum, which can be checked in several cases of interest.
 In these cases,  the true momentum (associated with translations in 
configuration space) is not fully conserved, because application of a 
magnetic field spoils translation invariance in the Klein-Gordon operator, 
but it turns out that the so-called  {\it twisted translations} [2]
 operating in phase space,  preserve the dynamics,
 which results in the occurence of a conserved vector [2-\dzy ]
 which reduces to ordinary momentum in the no-field limit.

At least this statement is true for $n$ charges in the Galilean theory,
 supposing  that the  mutual interaction between two charges corresponds 
to  a central force [2].

It is also true for one-body and two-body relativistic systems, under rather 
general assumptions.
Checking the one-body case is straightforward: the four-velocity being 
 $p - e A $, and the gauge being 
$ A = {1 \over 2} q \cdot F  $, it is a mere exercise to verify that 
$ C = p + e A $ commutes with the Klein-Gordon operator.

\medskip
 
For scalar particles, the {\it two-body problem}
can be posed in terms of a pair of   coupled wave equations [\rep ,\coupl ]. 

  In the absence of external field, the dynamics of the system has all the  
Poincar\'e symmetries, in particular it is invariant under ordinary 
translations; as a result total momentum $P$ is conserved.
Then, imposing sharp values to its components and after elimination of 
the relative time, one is left with a reduced wave equation involving three 
degrees of freedom only.
   At least when the mutual interaction potential does not depend on the total 
energy [\sazjmp ], this reduced equation can be seen as a standard  eigenvalue 
problem (in terms of the energy of relative motion).
As the constituent masses have been fixed from the outset, solving this 
equation yields a  spectrum (generally with a discrete part).
Admissible  values of  $P^2$ are correspondingly selected.

\medskip       
 
When an external electromagnetic field is applied to the system, 
a first problem consists in keeping the wave equations  consistent with one 
another.     When the  field can be seen as purely magnetic and constant
 (in space and time) for some observer, the compatibility requirement,
 combined with symmetry conditions and the demand 
of reasonable limits when mutual (resp. external)  coupling is turned off, 
is satisfied in closed form by an ansatz [1][\bij89 ][\dv95 ],
 provided the  that the  
mutual interaction term is known in the isolated system.

   Now the customary momentum $P$ is not any more a constant of the motion.
But we have previously proved this result [1,\dv95 ]:
 for a large class of mutual interactions 
(sufficient for all pratical purposes) 
 the {\it total pseudomomentum  of (globally) neutral systems}
 (that is $e_1 = -e_2 = e $)            is actually conserved.
 Moreover the four components of this vector commute among themselves.
This result seems to admit an extension to the case of an arbitrary constant 
electromagnetic field [\dv95 ],
 but only the pure electric or magnetic cases lead to a 
rigorous statement and to wave equations written in closed form, after a 
suitable transformation [\purmagn].

It is possible to take advantage of pseudomomentum conservation in order to 
reduce the degrees of freedom, in a way which parallels the customary 
separation of center-of-mass variables usually performed in isolated systems.
In addition, a further degree of freedom can be separated out.  
{\it  In the magnetic case} (considered throughout this paper),
 this degree of freedom is a timelike variable identified as  relative time 
in one of the possible lab frames.

    This situation provides a description of neutral
bound states in a magnetic  field.
Indeed all the components of pseudomomentum can be simultaneously 
diagonalized. In a generalized sense, the motion of the system as a whole is 
separated out, and inner motion can be considered; in the same spirit as 
in the nonrelativistic theory, it is natural to associate a spectrum to this 
(pseudo) relative motion.
  
   Solving the reduced eigenvalue equation will   determine 
 admissible values  for the square of pseudomomentum;
having bound states in mind, 
we shall be more specially interested in the discrete spectrum.

       Since pseudomomentum is the generalization of momentum in the 
presence of  field,  its square plays the role of an {\it effective squared 
mass\/} and should be observable.
  We are led to  investigate  how much  this quantity is shifted by the field
from the total squared mass we would have obtained for the isolated system.

\medskip

In a suitable representation, we shall explicitly write down and reduce the 
 wave equations.  

  As the reduced equation is a three-dimensional problem, it is tempting to 
apply the standard methods of perturbation theory.
But, even in the simple case where the unperturbed motion is ruled by a mutual
 interaction term which does not depend on $P^2$, a complication arises due to 
the presence of external field:
 the reduced equation nonlinearly depends on the total energy.
As a result, it is not an eigenvalue equation in the usual sense.
Fortunately, a spectral theory for {\it energy-dependent perturbations of a 
standard eigenvalue equation\/}     has been 
already presented in the literature [\rst ].

\bigskip
 
Let us now say a few words about the geometric elements corresponding to 
 a  constant magnetic field in the presence of  a constant timelike vector 
   $k$,     eigenvalue of pseudomomentum.

A  constant magnetic field  $F^{\mu \nu} $ provides a {\it unique and
 relativistically invariant} 
decomposition [1]  of any four-vector $\xi $  into  
longitudinal and transverse parts, $\xi = \xi \sl + \xi \st $.

In  particular 
$ k = k\sl + k \st $. Assuming that $k$ is timelike, it is clear that $k\st $ 
can accidentally vanish whereas $k\sl$ never does. In fact $k\sl $ is always 
timelike. Since $k\sl  \in (E\sl) $ it can be considered as defining a 
preferred  lab frame, we shall refer to it as the {\it special lab  frame\/}.
Let us emphasize that this  preferred lab frame also 
depends on the state of motion of the system (in contradistinction to the 
unique lab frame associated with an inhomogeneous field).

It is noteworthy that $C \sl ^\alpha =P \sl ^\alpha $ and therefore 
$ |k \sl | $ is nothing but the energy seen in the special lab frame.

\medskip
 
 In general the frame adapted to $k$ (which differs from $k \sl $) 
cannot be a lab frame.  In view of results about pseudomomentum in the 
Gallilean theory, it is reasonable to admit that $k$ carries 
information about the motion of the system as a whole.  So we call 
  {\it pseudo-rest frame} the frame adapted to $k$.

With respect to the special lab frame,  it has a velocity characterized  by the 
  {\it motional  parameter \/} 
$ \displaystyle     \epsilon  = { |k \st | \over | k \sl | } $.
In the particular case where $k\st = 0$, then $k \sl $ and $k$ coincide and 
 $\epsilon $ vanishes; there exists a frame adapted to $k$ where $F $ is 
purely magnetic and the subsequent developments get drastically simplified.
Otherwize, terms involving the contraction   $k \cdot F  $ arise. We notice 
that 
$ (k \cdot F)^\alpha = |k| E ^\alpha $ where $E^\alpha$ is the electric field 
"seen"  by an observer moving with momentum $k^\alpha $ ({\it motional electric 
field}).

\medskip
  This paper is organized as follows. In next section we recall the 
ansatz which provides explicit equations of motion, and  simplify
a  quantity $\widehat Z$ which was  an essential ingredient in  the 
formulas of refs.[1, \dv95 ].
In Section 3, several  expressions given in compact form in that previous work 
are explicitly developed for applications.
Section  4 is devoted to the three-dimensional reduction, with emphasis on the 
role of relative energy.  Various terms present in the reduced equation are 
discussed and ordered according to the powers of field strenght and motional 
parameter.
In Section 5 we consider  normalization and
 the possibility of a reliable perturbation treatment.
Last Section is devoted to a few concluding remarks.

\bigskip
 
{\bf 2. BASIC EQUATIONS}

A system of scalar particles can be described by a pair of coupled Klein-
Gordon equations
$ \displaystyle 
H_a   \Psi = {1 \over 2} m _a ^ 2  \  \Psi  \qquad \quad  a,b = 1,2     $
where $\Psi $ has two arguments $q_1, q_2$ running in spacetime.
We  cover all cases of practical interest assuming that 
$$ H _a = K_a + V $$
 In this formula  $2K_a$ is the squared-mass operator 
for particle $a$ alone in the magnetic field, and $V$ is a suitable 
modification of the term  $ V \zer $ which describes the mutual interaction 
in the absence of external field; more generally, the label $(0)$ refers to 
the no-field limit of any quantity.

In order to be more specific we separate canonical variables in two classes 
as follows.
$$ P= p_1 + p_2,  \qquad \quad Q = {1 \over 2} (q_1 + q_2)   $$
$$ z =q_1 -q_2,  \qquad \quad  y= {1 \over 2} (p_1 - p_2)   $$
This provides two sets of standard commutation relations.
Useful quantities are 
$$ {\widetilde y}^2 = y^2 - (y \cdot P )^2 / P^2 $$
$$ {\widetilde z}^2 = z^2 - (z \cdot P )^2 / P^2 $$
The latter is an essential ingredient of mutual interactions; but in order to 
avoid denominators in calculations, it is  convenient to  employ 
  $$  Z = z^ 2  P ^ 2  -  ( z \cdot  P )^ 2    \eqno (2.1)  $$
  We shall  assume that 
      $$  V\zer = f(Z, P^ 2, y \cdot  P)   \eqno (2.2)$$ 
Since our system is globally neutral, it turns out that pseudomomentum is 
$$ C = P + {e \over 2 } z \cdot F    $$
In canonical variables we can separate the transverse pieces from the 
longitudinal ones, for instance $P = P\st + P \sl $, etc.
Transverse and  longitudinal variables mutually commute.

      Compatibility requires that
 $  [    K_1 - K_2,  V   ] = 0    $.
Adapting Bijtebier's method [\bij89 ]
 to the peculiarities of the constant magnetic 
field, we have performed  a  canonical  transformation  say
$$  \Psi ^\prime = \exp (iB) \  \Psi,
\qquad \quad {\cal O}  ^\prime  =
  \exp (iB) \ {\cal O} \ \exp (-iB)    \eqno (2.4) $$   
where   $ {\cal O} $ is any operator.
    $B $ was choosen [1,\rigo ] such that transformation (2.4) yields 
$$  K'_1 - K'_2 =  y\sl \cdot P \sl           \eqno (2.5)          $$
thus compatibility is satisfied through the ansatz [1]
$$    V^\prime  =  f( \widehat Z, P^2, y\sl.P \sl )        \eqno (2.6)  $$
where 
$ \widehat Z= Z' \zero=    (Z')_   {F=0}     $ (it turns out that 
$     \widehat Z $ commutes with $  y\sl \cdot P \sl $).
The explicit form of $ \widehat Z $ was calculated in ref.[1]
      $$ \widehat Z =
Z  +  2(P\sl^2 \   z \cdot P - P^2  \   z\sl \cdot P\sl  ) L
  +  P\st^2 P\sl^2  L^2                    \eqno     (2.7)  $$
   where the scalar  $L$ is defined as 
       $   L  =  L \cdot  z $  in terms of the  the four-vector [\L.L ] 
 $$  L^ \alpha = {P \sl ^ \alpha  \over  (P \sl )^ 2}   \eqno (2.8) $$
As it stands, formula (2.7) is of poor  practical interest. 
It is essential to observe [1] that 
 $ \widehat Z $ commutes with $y\sl \cdot P \sl $.
 Let us transform (2.7) in order to render this  property manifest.
First we split  $z$ as the sum of $z\sl $ and $z \st $ in   $Z$, hence
$$ Z = (z\st ^ 2 + z \sl ^ 2 ) P ^2  
- (z \st \cdot P )^ 2 - ( z\sl \cdot P )^ 2
-2 (z \st \cdot P )(z \sl \cdot P )            \eqno (2.9)      $$
Develop (2.7) and perform  elementary manipulations using (2.9). We get
$$ \widehat Z =
Z  +  2 P\sl ^ 2 (z \st \cdot P ) L  
+ 2 P \sl ^ 2 (z \sl \cdot P ) L
-2 {P^2 \over P \sl ^ 2} (z \sl  \cdot P)^ 2
+ {P \st ^ 2 \over P \sl ^ 2 } (z \cdot P \sl )^ 2     $$
Using (2.9) again we obtain
$$ \widehat Z =    Z
+ 2 (z \st \cdot P )(z \sl \cdot P)    
+2 (z \sl \cdot P )^2  - 2 {P ^ 2 \over P \sl ^ 2} (z \sl \cdot P )^ 2
+ {P \st ^ 2 \over P \sl ^ 2} (z \sl \cdot P )^ 2    $$
That is 
$$ \widehat Z = Z + 2 (z \st \cdot P )(z \sl \cdot P) 
-(z \sl \cdot P)^ 2   {P \st ^ 2 \over P \sl ^ 2}            $$
Using (2.9) we notice cancellation of the terms proportional to 
 $     (z \st \cdot P )  (z \sl \cdot P)$ and we  can write 
$$ \widehat Z = z \st ^ 2 P^ 2 + z \sl ^ 2 P^ 2 - (z \st \cdot P)^ 2
- (z \sl \cdot P)^ 2 (1 +  {P \st ^ 2 \over P \sl ^ 2} )               $$
That is to say
  $$ \widehat Z =  
 z \st ^ 2 P ^ 2 - (z \st \cdot P)^ 2  + 
 P^ 2 (  z \sl ^ 2   -   {  (z \sl \cdot P \sl ) ^ 2 \over P \sl ^ 2}  )
     \eqno (2.10) $$

It is convenient to define  the projector  "orthogonal" to $P \sl$, say
$$ \Omega ^ \alpha _ \beta =   \delta ^ \alpha _\beta  - { P \sl ^ \alpha
 { P \sl }_\beta   \over P\sl ^ 2 }     \eqno (2.11)  $$
 because we can write
$$  z \sl ^ 2    -  { (z \sl \cdot P \sl ) ^ 2   \over P \sl ^ 2}  
 =  (\Omega z \sl )^ 2                                        \eqno (2.12) $$
and we easily check that  $ (\Omega  z) ^\alpha $  commutes with    
    $ (y \sl \cdot P\sl  )$.
So we finally have
$$ \widehat Z = z \st ^ 2  P^ 2 - (z\st \cdot P)^ 2 
   +  (\Omega z \sl )^ 2   P^ 2                       \eqno (2.13)    $$
which justifies the claim that   $ \widehat Z$  
commutes with   $y\sl \cdot P\sl$
Here we notice that 
     $ \Omega z \st  = z \st $ and  finally obtain
$$ \widehat Z   = (\Omega z)^ 2 P^ 2 - (z \st \cdot P )^ 2    \eqno(2.14) $$
This simplification of (2.7) renders the ansatz more tractable.

  Remark:
Formula (2.14) was derived without specifying the respective dimensions of 
the longitudinal and transverse spaces.
It is valid also in the case considered in ref.[\bij89 ],
 but in that case $\Omega z $ 
reduces to $z \st $, which makes  $ \Omega z \sl $ to vanish.

\medskip
 
It is convenient to replace the basic wave equations by their sum and 
difference,  setting 
$$ \mu = {1 \over 2}(m_1 ^2 + m_2 ^2), 
\qquad \quad \nu = {1 \over 2}(m_1 ^2 - m_2 ^2 )$$
After transformation (2.4), and in view of (2.6)  we need the expression of 
 $K'_1 + K '_2$.  Equations  (3.36) of Ref.[1]  yield in the present notation
  $$ K'_1 + K'_2 =
K_1 + K_2  -2T (L \cdot y)  + T^2 ( L \cdot L)   \eqno (2.15)  $$      
where 
 $$ T = K_1 -K_2 -  y\sl \cdot P\sl                             \eqno(2.16) $$
and the four-vector $L$ is given by (2.8).
We notice that
$$   L \cdot L = {1 \over  ( P \sl ) ^ 2} ,    
   \qquad  \qquad 
  L \cdot y  =  {y \sl \cdot P \sl  \over  P \sl  ^2 }     \eqno (2.17) $$

\medskip
 
For neutral systems, a further transformation  inspired by the work of Grotch 
and Hegstrom [4]  permits to get rid of the $Q$ variables.
We introduce a new wave function 
$ \Psi '' = ( \exp i\Gamma) \Psi ' $ 
with the help of  the unitary transformation generated by 
  $$   \Gamma  =  {e \over 2} (z.F.Q)         \eqno  (2.18)        $$
We  set 
  $$ {\cal O} ^\sharp  = \exp (i \Gamma) \ {\cal O}  \   \exp (-i \Gamma) 
  \qquad      {\cal O} '' =
 ({\cal O}' ) ^ \sharp    \qquad  \forall  {\cal O}         \eqno (2.19)   $$

By (2.19), pseudomomentum is transformed to $P ^ \alpha $.
Since $y \sl \cdot P \sl $ and  $ P\sl ^ \alpha $ commute with $ \Gamma $,
they are  not affected by transformation (2.19), thus  finally
$$ H''_1 - H''_2 = K''_1 -K''_2 =   y\sl \cdot P \sl   \eqno (2.20)  $$
Taking (2.20) into account, consider sum and difference of the wave equations. 
$$  ( H''_1 + H''_2 ) \Psi ''   =   \mu \Psi ''     \eqno (2.21)         $$  
 $$  y\sl \cdot  k \sl \  \Psi'' =  \nu  \Psi ''          \eqno  (2.22)     $$
Transformation (2.19) ensures that  $ H''_1 + H''_2 $ doesnot depend on the 
external variable $Q$. Moreover    $ H''_1 + H''_2 $ commutes with 
$y \sl  \cdot P\sl $;  therefore  it should not involve 
 the operator $ z\sl \cdot P\sl $ conjugate to it. We shall explicitly check 
this point in Section 4.

\medskip
    From now on 
 {\it we demand that $\Psi ''$ be eigenstate of pseudomomentum with a 
timelike four-vector   $k^ \alpha $} as eigenvalue. 
Combining this requirement with (1.22)   we obtain    
$$  \Psi'' = \exp (i k \cdot Q) 
\ \exp (i \nu { z\sl \cdot k \sl \over  | k \sl | } )\  
 \phi             \eqno (2.23)   $$  
where  $ \phi  $ depends   on $z$,  but only through its 
projection orthogonal to $k \sl $, and additionally depends on $k$ and on 
$\nu$ as parameters. 
In other words $ \phi = \phi ( \nu ,k,  \varpi  z) $ where we define 
$\varpi $ as the tensor which projects any vector on the 3-plane orthogonal to 
$k \sl $.
The linear space of such functions includes  as a subspace  the Hilbert space 
$$  L^2 (k\sl)  =  L^2 ({\bf R}^3, d^3 \varpi z)       \eqno (2.24) $$
 characterized by convergence of the triple 
integral  $\int \phi ^* \phi  d^3  \varpi z$ which simply reads
$ \int \phi ^*  \phi  d^3 {\bf z} $  in any frame adapted to  $k \sl$.
We use the subscript $k$ for typographical simplicity; it refers to the {\it 
vector \/}    $k$. In fact, the Hilbert space defined above  depends on $k$ 
only through  the {\it direction \/} of $k \sl$
 (it doesnot depend on $k\sl ^2 $).

   Remark: There is no explicit contribution of  external field
to  equations (2.22) (2.23).  In contradistinction, 
 equation (2.21) remains  sensitive to the presence of external field.

\bigskip
 
{\bf 3.   EXPLICIT FORMULAS}

Our goal is to discuss the system (2.21)(2.22). In order to have an explicit  
expression for (2.21), let us first calculate   $ H''_1 + H''_2 $.
 According to the notation set in (2.19) and to the definition of $\Psi ''$ we 
can write
 $$   H''_1  + H''_2 =  (K'_1 + K'_2) ^ \sharp  +  2V'^ \sharp 
=  K''_1 + K''_2 + 2V''  \eqno (3.1) $$ 
We have  to   transform  $ (K'_1 + K'_2) $ and $ V'$ as indicated in 
(2.19). But  $V'$ is given by  (2.6).
  According to (3.1),  we need to transform  formulas  (2.15) and (2.6).
Notice that the transformation generated by $\Gamma $ 
always can be carried out explicitly.
It leaves  $ q_1 , q_2 $ unchanged whereas
  $$   p_1 ^ \sharp  = p_1  - {e \over 2}   F  \cdot q_2   \qquad \   \quad
   p_2 ^ \sharp  = p_2  + {e \over 2}   F  \cdot q_1  \eqno (3.2)  $$
Hence 
$$ P ^ \sharp = P  + {e \over 2}  F \cdot z  \eqno (3.3) $$   
  $$ y^ \sharp = y  - {e \over 2}  F  \cdot  Q     \eqno (3.4) $$
Notice that  $ F \cdot z $ 
has only transverse components and they depend on the 
transverse variables   $z\st $ only.
Therefore we can write 
$$  P \sl ^ \sharp  =  P\sl \eqno (3.5) $$
Eq (3.3) also implies
$$ P^ {\sharp 2} = P^2 + e P \cdot F \cdot z
+ {e^2 \over 4} (F \cdot z)^2                      \eqno (3.6)  $$
$$  (P \st ^ \sharp )^2  =
P \st ^2  +  e (P \cdot   F \cdot  z)
+  {e^2  \over 4}  (F \cdot z) ^2              \eqno  (3.7)       $$
These formulas will be useful later on.

Straightforward calculation made in ref [1]  yields 
$$  K_a ^ \sharp =   {1\over 2} (p_a - {e \over 2}  z \cdot  F ) ^ 2   \eqno 
(3.8)  $$
Obviously  $ z ^ \sharp =  z $,  and $ P ^ \sharp $ was given in (3.3) above.
Explicit development gives
$$ 2K_1 ^ \sharp =
p_1^ 2  -  {e \over 2} p_{1 \alpha} z_\sigma  F^ {\sigma \alpha} 
 - {e \over 2} z_\sigma F ^ {\sigma \alpha}  p_{1 \alpha}
+  {e^ 2 \over 4}  ( z \cdot F) ^ 2      \eqno   (3.9)                  $$   
$$ 2K_2 ^ \sharp =
p_2^ 2  -  {e \over 2} p_{2 \alpha} z_\sigma  F^ {\sigma \alpha} 
 - {e \over 2} z_\sigma F ^ {\sigma \alpha}  p_{2 \alpha}
+  {e^ 2 \over 4}  ( z \cdot F) ^ 2              \eqno (3.10)          $$  
We observe that all terms quadratic in the charge disappear from 
the difference; we get
$$ 2(K_1 - K_2) ^ \sharp =
p_1^ 2 - p_2^ 2  -e ( y_\alpha z_\sigma    +  z_\sigma y_ \alpha) F^ {\sigma 
\alpha}  $$
As  $ [    z_\sigma , y_ \alpha   ] $ is symmetric as a tensor,  the   order 
of $y, z$ is  immaterial when multiplied by the skew-symmetric $F$. Recall that 
$ p_1 ^2 - p_2 ^ 2 = y \cdot P $, hence  the useful formula
$$  (K_1 - K_2 )^ \sharp =  
y \cdot P -  2 e \    z \cdot F  \cdot y       \eqno (3.11)      $$

\medskip
 
But our goal was  to   transform  eq (2.15). We first calculate 
 $  ( K_1 + K_2) ^ \sharp $.

Notice that 
 $ P \cdot ( z \cdot F ) =  (z \cdot F ) \cdot  P  = 
   z \cdot  F  \cdot  P   $, hence 
$$       2  ( K_1 + K_2) ^ \sharp =  p_1 ^ 2 + p_2 ^ 2  
- e   z \cdot  F  \cdot P   +  {e^ 2 \over 2 }  (z \cdot  F) ^ 2        $$
But
 $ \displaystyle   p_1 ^ 2 + p_2 ^ 2  \equiv 
            P^ 2 / 2  + 2 y^ 2  $                        thus  
$$  ( K_1 + K_2) ^ \sharp    =
{P^ 2 \over 4 }  + y^ 2   - {e \over 2}   z \cdot F  \cdot P   
+ {e^ 2 \over 4}  (z \cdot F ) ^ 2                          \eqno (3.12)    $$
Then we have to transform  $T,  \  L \cdot y $  and $L \cdot L $.
Take  (2.16) into account and remember (3.11).
 (we know that   
$ y \sl \cdot P\sl $ and $P\sl $ are not affected by $\sharp$). We get
$$ T ^ \sharp =
y \st \cdot P \st - 2e z \cdot F \cdot y    \eqno (3.13)     $$
And finally after a glance at (2.17) we observe that
  $ L \cdot y $ and $ L \cdot L $  are unchanged in the transformation 
generated by $\Gamma $.

Now we apply  transformation (2.19) to (2.15),
taking (3.12)(3.13) into account.  It gives 
 $$  K''_1 +  K'' _2 =      $$
$$          {P^ 2 \over 4} + y^ 2   -  {e\over 2}  z \cdot F  \cdot P 
 + {e^ 2 \over 4}  (z \cdot F )^ 2
- 2 ( y\st \cdot P\st  -2e  z \cdot F \cdot y ) L \cdot y
+  (y\st \cdot P\st - 2e  z \cdot F \cdot y )^ 2   \  L \cdot L  
\eqno  (3.14)      $$
We know that $2 V''$ must be added to this expression in order to obtain
 $H''_1 + H''_2  $.
  But in (2.18)             $F ^ {\mu \nu} $ is 
purely transversal, therefore 
   $ (y \sl \cdot P \sl)^ \sharp  = y \cdot P\sl    $.
We have by  (2.6)
$$ V'' = f( {\widehat Z} ^ \sharp, 
 {P^\sharp }^2,    y \sl \cdot P\sl )   \eqno (3.15)  $$
where $ {P^\sharp }^2 $ is as in (3.6) and we must compute
 $\widehat Z ^ \sharp $ from (2.13) by help of (3.3)(3.5).
We make the convention that 
  $\widehat Z ^ \sharp   = ( \widehat Z) ^ \sharp $ (and not the reverse).

\medskip
 
Then   we apply the transformation (2.19) to eq. (2.13). Inspection of (3.5) 
shows that  $ (\Omega z \sl )^ 2 $ is not affected by the transformation.
We notice that  
$ z \st \cdot P ^ \sharp =  z \st \cdot P $ because, $F$ being purely 
transverse, 
$ z \st \cdot  F  \cdot z $ identically vanishes.
Thus
$$ \widehat Z ^ \sharp =
{P^ \sharp}^ 2  (\Omega z )^ 2- (z \st  \cdot  P )^ 2     \eqno (3.16)  $$
Now, eqs (3.1)(3.14)(3.15) supplemented with (3.6)and (3.16) furnish
the complete expression of  $ H''_1 + H''_2 $.

\bigskip 
    {\bf 4. THREE-DIMENSIONAL REDUCTION}

{\bf 4.1  The reduced wave equation for isolated systems}

Our presentation of the three-dimensional reduction will be more transparent 
if we first remind how it is usually carried out (and how it leads to a 
spectrum for $P^2$) when the system is free of external forces.

In this subsection we drop the superscripts $(0)$ referring to an isolated
 system.
In the absence of external field, the squared-mass operators  can be written  
    $ \displaystyle  2  H_a =    p^2 _a + 2 V     $.
A little of elementary algebra provides 
$$   H_1 + H_2 =    {P^2 \over 4} + y^2 + 2 V                  \eqno (4.1)  $$
$$   H_1 - H_2  =  y \cdot  P                                  \eqno (4.2)  $$
and the wave equations take on the form
$$    ( {P^2 \over 4} + y^2 + 2 V )\Psi =  \mu \Psi      \eqno (4.3)  $$
$$    ( y \cdot  P  ) \Psi =   \nu  \Psi                  \eqno (4.4) $$
Introducing  
$   \widetilde y  ^2    $ as defined in Section 2           we get  
$$    H_1 + H_2  =
{P^2 \over 4 } +  {  (y \cdot P ) ^2   \over  P^2  }   
+   \widetilde y  ^2   +  2V                           \eqno (4.5)   $$        
and the wave equation (4.3) now reads
$$   ( {P^2 \over 4 } +  {  (y \cdot P ) ^2   \over  P^2  }   
+   \widetilde y  ^2   +  2V  )\Psi =   \mu  \Psi        \eqno (4.6)       $$
Introduce the operator [\rep , \anal ] 
$$  N =        \widetilde y  ^2   +  2V                  \eqno   (4.7)    $$
 equation (4.6) takes on the form
$$  ( {P^2 \over 4 } +  {  (y \cdot P ) ^2   \over  P^2  }
    + N   )\Psi  =    \mu  \Psi                                \eqno (4.8)   $$
It is noteworthy that    $-N $ is intimately related with  the energy of
 relative   motion (see the nonrelativistic limit $\lambda \ll \mu$). 
Taking  (4.1)(4.2) into account it is easy to check the identity
$$ -N   = 
{P ^2  \over 4} +  {(H_1 - H_2)^2  \over P ^2}
                                          -  (H_1 + H_2)   \eqno (4.9)    $$
Let us assume that $\Psi $ is also eigenstate of the linear momentum, that is 
$ P ^\alpha \Psi =   k^\alpha  \Psi $, where  $k$ is timelike.

   Since $P^2$ and $y \cdot P$ are simultaneously diagonalized, so are the 
quantities $P \cdot p_1$ and $P \cdot  p_2$. 
It is usually required  that the eigenvalues of these operators are 
both positive, which amounts to say 
$$             \half  k^2   >  |\nu |           \eqno  (4.10)  $$
Justification for this condition is easy when   the mutual interaction is
$ V = f (Z, P^2) $. In the underlying classical theory [\anal ]
 the equations of motion imply that 
$   \displaystyle
  P \cdot z  =   P \cdot p_1   \    \tau _1 - 
                   P \cdot p_2   \   \tau _2 + const.      $
where  $\tau _1 , \tau _2$  are the evolution parameters (generalizations of 
the proper times).
On the equal-time surface  $(\Sigma ) \simeq  P \cdot z = 0 $ we get the 
relation
$ \displaystyle         P \cdot p_1 \   \tau _1 = 
          P \cdot p_2 \   \tau _2 + const.         \     $
between the parameters. It is clear that $\tau_1$ must increase together with 
$\tau _2$, which implies that {\it classically\/} $P \cdot p_1$ and $P \cdot 
p_2$ have the same sign (necessarily positive, since their sum is $P^2$).
Equation (4.10) is just the quantum conterpart of this condition. 
Extension of this argument to more general interactions is an open question, 
but  other  considerations motivated by the Bethe-Salpeter equation and 
propagator theory [\sazdj]  strongly support condition (4.10).

\medskip
 
We can obviously write
$$ \Psi = \exp (i k \cdot Q) \psi (z)           \eqno (4.11)    $$
Using also (4.4) we cast eq. (4.8) into the form 
$$ ( {k^2 \over 4 } + { \nu ^2 \over k^2 } + N)\Psi    
=   \mu \Psi                                           \eqno (4.12)     $$    
Let us set 
$$ \lambda =  
        {k^2 \over 4 } + { \nu ^2 \over k^2 } - \mu         \eqno  (4.13)  $$
(4.12) now reads 
$$  (N + \lambda )\Psi = 0      \eqno (4.14)         $$
which suggests to consider (4.14) as an eigenvalue problem where  $ \lambda $ 
is the eigenvalue for the operator $-N$.
Caution is needed however, in the cases where $V$ actually depends on $P^2$ 
({\it energy-dependent case\/});
 in this case $k^2$ may be substituted to $P^2$ into
the potential and further eliminated from (4.14) 
with help of (4.13) (as we shall see there is only one way to solve (4.13) for 
$k^2$). But one is left with an extra  dependence on $\lambda $ in 
equation  (4.14).
This complication, first pointed out by Rizov, Sazdjian and Todorov [\rst ],
 will be discussed below, in the context of the reduced 
problem.

\medskip
           Indeed equation (4.4) implies that the form (4.11) of the
 wave function can be further reduced.  Recall that   
$\displaystyle 
  y_\alpha  =   -i   \partial  /   \partial  z ^\alpha       $.
In a frame adapted to $k$, we can write   $ y \cdot k =   y _0  k^0 $   
thus (4.4) entails 
$  \psi  =  \exp  (  i \nu z^0 / k ^0) \    \phi          $
where $\phi $ doesnot depend on $z^0$.                                       
In fact $\phi $ belongs to a linear space which depends on the direction of 
the vector $k^\alpha $, but {\it doesnot depend \/} on the number 
 $k = \sqrt {k^\alpha k_ \alpha } $. Imposing  square integrability defines
a Hilbert space, say   $L^2 ({k^\alpha}) $.
Now eq. (4.8)  or (4.12) becomes  an equation to be solved in
 $L^2 ({k^\alpha}) $            because after substitution of 
$k$ for $P$ and  $\nu $ for $ (y \cdot  P) $ in $N $ we obtain an operator 
acting  in $ L^2 (k^\alpha) $, say  $(N) _{\nu , k} $.
  Insofar as $V$ depends on $P^2$, the operator  $(N) _{\nu , k} $  
 depends on the scalar $k^2$.    
We can solve (4.13) for $k$ and substitute the result in  $ (N)_{\nu , k} $.
In other words, we insert 
$$ k^2 =    2  (\lambda + \mu ) +
  2  \sqrt {(\lambda + \mu )^2  - \nu ^2}                    \eqno (4.15) $$
into  $ (N)_{\nu , k} $.

Notice that an alternative solution of (4.13) corresponding to the minus sign
 in front of  the radical is discarded because it would violate the condition
 $ \half  k^2   >  |\nu |  $.
In contradistinction, provided  that 
$ \displaystyle   \lambda + \mu   >  \nu  $  we are sure that this condition 
is satisfied by  solution (4.15). As a result the correspondance between 
$\lambda$ and $k^2 $ remains one to one for unequal masses.

   But, for energy-dependent interactions, 
we end up with a reduced wave equation which fails to be an eigenvalue 
equation in the usual sense, being  nonlinear in the eigenvalue  $\lambda$ 
[\klambda ].
This complication, pointed out in ref.[\rst ]
 stems from the possible dependence of the mutual interaction
 term $V$ on the (squared) total energy  $P^2$. 

             Fortunately, a mathematical 
theory exists for  generalized eigenvalue equations which are nonlinear in the 
eigenvalue [\kel ].
But this theory involves technical sophistications resorting to the 
concept of "associated vectors". Its  main departure from   the standard 
theory is that eigenvectors corresponding to different generalized eigenvalues
may fail to be orthogonal. This drawback (not characteristic of the two-body
problem, see [\wein ]) led the authors  of ref. [\rst ]
 to advocate a re-definition  of the scalar product of eigenvectors.
In this paper we  shall focus on  situations   where  this redefinition is 
unnecessary. 

\medskip
 
 No such problem arises when $V$ does not depend on the energy.
Any potential of the form 
$ V =  f (Z/P^2 , y \cdot P )$  
(in particular the harmonic oscillator  $V = const.   \widetilde z ^2 $)
 leads, in absence of external field,
 to an ordinary eigenvalue equation in 
$ \displaystyle  L^2 (k^\alpha )  = L^2 ( {\bf R} ^3 ,  d^3  z _ \perp ) $, 
where $z_\perp $ is orthogonal to $k$.

   This class of potential encompasses a lot of phenomenological models.
In contrast, realistic potentials motivated by QED pertain to  the other type.

\bigskip
                 
{\bf 4.2 The reduced equation in the presence of external field}.

Let us now return to the case where an external potential is applied to our 
system.   The aim of this section is only to reduce the wave equations.
The discussion of all matters concerning the spectrum are postponed 
to Section 5.

Since $C$ is the natural generalization of linear momentum in the presence of 
external field, and by analogy with (4.9),  we shall {\it define \/}
$$ -N   = 
{C ^2  \over 4} +  {(H_1 - H_2)^2  \over C ^2} 
                                         -  (H_1 + H_2)   \eqno (4.16)   $$
Although formula  (4.7) is no longer valid, the sum of the basic
 wave equations still  reads
$   (N + \lambda ) \Psi  = 0 $  with $\lambda $ defined as in (4.13).
We assume that $\Psi $ is eigenstate of the pseudomomentum, say 
$ C ^\alpha \Psi = k ^\alpha \Psi$, for some constant $k$ which is supposed to 
be strictly {\it timelike}.
By analogy with the isolated system and also for pragmatic reasons, we shall 
confine ourselves to the sector $ k^2 / 2  >  |\nu |$. 
After transformation to the convenient representation we get
$$-N'' =
{P ^2  \over 4} +  {(H''_1 - H''_2) ^2   \over P ^2} 
  -  (H''_1 + H''_2)                                          \eqno (4.17)    $$
and our goal is to determine the spectrum of this quantity. Then one will
pass from $\lambda$ to $k^2$ through (4.13).

\medskip
  The calculations can be organized as follows:

  Whereas (2.22) 
fixes the dependence in the relative time, eq.(2.23) allows us 
to factorize out the "center-of-mass motion", and we are left with the reduced 
wave function    $ \phi$     which arises in eq. (2.23).
Obviously  (2.22) implies that  
$$ y \sl \cdot k \sl  \  \phi  = 0         \eqno (4.18)  $$
thus  $\phi$  depends on $z $ only through its projection $ \varpi z$.
Imposing square integrability amounts to  require  that $\phi$ is in the 
Hilbert space $L^2 _k $ defined by (2.24). It is clear that $\phi$ generally 
depends  on $ \nu $ and $ k$ as parmeters.

In order to write down the  reduced wave equation, we  replace $P^ \alpha$ 
 and  $  y \sl  \cdot P \sl $     respectively 
 by  their  eigenvalues  $k^ \alpha $     and  $ \nu $ 
in $H''_1 + H''_2 $,
 and we divide  by exponential factors.

For any operator ${\cal O} $ it is convenient to use the following convention
$$    ({ \cal O})_{\nu , k } =
{\cal O}   |   _{ y\sl \cdot P \sl = \nu , \    P=k }       
\eqno  (4.19)  $$
The subscript $k$ refers to the vector $k$, which finally contributes by its 
longitudinal piece only.
Notice that here a term like $y^2$  must be  written  as
  $ \displaystyle 
y^2 \equiv  (\Omega y  )^2 +  { ( y \sl  \cdot  P \sl )^2  \over P \sl ^2 }  $.
  If we now introduce the projector $\varpi $ orthogonal to $k \sl $, 
and  use  the identity
 $ \displaystyle {1 \over k \sl ^2 } = {1 \over k ^2} (1 - \epsilon ^2 ) $     
we obtain 
$$   ({P^2 \over 4 } + y ^2 ) _{\nu, k} =
  { k  ^2 \over 4} + (\varpi y)^2  + {\nu ^2  \over k \sl ^2 }     
=    { k  ^2 \over 4} + (\varpi y)^2  + {\nu ^2  \over k^2} 
 - \epsilon ^2 {\nu ^2 \over k ^2 }                \eqno  (4.20)   $$ 
 which is to take into account when computing  
 $ (K''_1 + K''_2) _ {\nu , k } $ from (3.14).

Naturally  
$ \displaystyle    ( H''_1 + H ''_2)  _{ \nu , k  } = 
(K''_1 + K'' _2 )_{\nu , k }  +
     2  (V'') _{\nu , k   } $.  
Defining
 $$  R (\nu, k \sl, k \st) =   (K''_1 + K''_2)_{\nu, k  }   \eqno (4.21)    $$
$$ W (\nu, k \sl, k \st) =     (V'')_{\nu,  k }      \eqno (4.24)   $$
we end up with the equation
$$     R    \    \phi  +    2W   \  \phi  =  \mu \phi    \eqno (4.23)  $$
At this stage, let us observe  that, in agreement with a remark made in 
Section 2,
 it is expected that  neither   $R$ nor  $W$  involve 
 $z\sl  \cdot P \sl $. This will be  checked later on and will permit us to 
consider $R$ and $W$ as operators acting in  the Hilbert space
 $L^2 (k \sl) $ defined by  (2.24).

   We remind that $\mu $ is just a parameter  fixed from the outset.
The question wether (4.23) can be considered as a spectral problem,
 {\it and for which eigenvalue\/}, will be considered later on, with help of 
equations (4.13)(4.16). See eq. (4.33) below.

\medskip

The explicit expression of  $R$ comes from (3.14),  with help of (4.21),
and at first sight     seems very involved.
This apparent complication results from having  used the necessary 
transformations  (2.4)(2.19).
Things become more tractable if we separate all terms involving
 $F ^ {\mu \nu}$ from those which survive when the field is turned off, and 
so on.    So let us perform such a separation in a systematic way.
Be cautioned that  $ ( {\cal O }'') \zer  \not=    ( {\cal O} \zer ) ''   $.

Since $K''_1,  K'' _2 $  are no more than quadratic in the field strenght, we 
can write
     $$ K'' _1  +  K''_2 = 
 (K'' _1 + K'' _2)  \zer  + (K''_1  + K'' _2)^    {(1)}    +
  (K''_1 + K '' _2) ^    {(2)}                          \eqno (4.24)     $$  
with the obvious convention that superscripts $(1),(2)$ respectively refer to 
(homogeneous)  linear and quadratic terms in the field strenght.
 Therefore
     $ R =  R \zer + R ^   {(1)} +   R  ^  {(2)}   $.  
In eq.(3.14) we have to 
replace $P$ by $k$ and $y \sl \cdot P \sl $ by $\nu $,  in order to compute 
$R$.      
 Remembering (2.17) we start from (3.14) and  compute 
$ K'' _1 + K'' _2 $ (to be inserted into (3.1) and further simplified by the 
above substitution).
We first consider the  piece  of it which survives in the no-field limit.
In eq. (3.14) we collect zeroth order terms in the field and find 
 $$  (K''_1 + K '' _2) \zer =  
  {P^2 \over 4 } +  { ( y \sl \cdot P \sl)^2 \over P \sl ^2}  +  
(\Omega y )^2   
+   y \st \cdot P \st  { y \st \cdot P \st - 2 y \sl \cdot P \sl   \over 
 P \sl ^2 }                                                \eqno (4.25) $$   
According to notation (4.19) we can write 
  $$  (K''_1 + K '' _2) \zer _ {\nu , k} =  
  {k^2 \over 4 } +  {  \nu ^2 \over k \sl ^2}  +  
(\varpi y )^2   
+   y \st \cdot k \st  { y \st \cdot k \st - 2 \nu   \over 
 k \sl ^2 }                                                  \eqno (4.26)  $$  
At this stage, one might be tempted to separate out the first two terms 
in the right-hand-side  because they  are constant.
However the second one  depends  on $k \sl $ which looses intrinsic meaning in 
the no-field limit.
Therefore we prefer to apply  the identity 
$ \displaystyle
 {1 \over k \sl ^2 } = {1 \over k ^2 } (1 - \epsilon ^2 ) $
which yields 
 $$  (K''_1 + K '' _2) \zer _ {\nu , k} =  
  {k^2 \over 4 } +  {  \nu ^2 \over k ^2}
 - \epsilon ^2 {\nu ^2  \over  k^2 } + (\varpi y ) ^2 
+ { (y \st \cdot k \st )^2   \over k \sl ^2 }
   -  2 \nu {y \st \cdot k\st   \over k \sl ^2 }       \eqno (4.27) $$
Now setting 
 $$ (S)_{\nu , k}  = (\varpi y )^2  + 
y \st \cdot k \st  { y \st \cdot k \st - 2  \nu    \over 
 k \sl ^2 }   
- \epsilon ^2  {\nu ^2  \over k^2 }   \eqno (4.28) $$ 
we can write
 $$  (K''_1 + K '' _2) \zer _ {\nu , k} =  
  {k^2 \over 4 } +  {  \nu ^2 \over k ^2} + (S)_ {\nu, k}       \eqno (4.29) $$
so we end up with 
$$ R \zer = 
 {k^2 \over 4} + { \nu ^2 \over k  ^2 } + (S)_{\nu , k }   \eqno (4.30) $$
The field-depending terms in (3.14) provide
   $$ R^ {(1)}=  4 e (z\cdot F \cdot y)    {\nu  \over  k \sl ^ 2 } 
         -  {e \over 2} z \cdot F \cdot k      \eqno (4.31)     $$ 
$$ R  ^ {(2)} =
 {e^ 2 \over 4} (z \cdot F)^ 2 
+  4 e ^ 2   { (z\cdot  F  \cdot  y)^ 2  \over k \sl ^ 2 }      
\eqno  (4.32)       $$
Remember that $F$ has transverse components only. Contractions involving $F$
only depend on the transverse components;  for instance  $F \cdot k $  is 
just a combination of the quantities  $k\st ^\alpha$.
It is noteworthy that only the transverse components of $z, y $ arise in 
$ \   R  ^{(1)}, \   R ^{(2)}  \   $, whereas $ (S)_ {\nu, k } $  depends on 
  $ \varpi y $ and   $ y \st $.
As a whole,  $R$ depend only on  $ \varpi z $ and  $ \varpi y  $ (recall $y 
\st , z \st $ are pieces of $\varpi y , \varpi z $ respectively).

Let us again define $\lambda $ by formula (4.13). 
In view of (4.30)(4.31)(4.32), equation (4.23) may be written
$$  \lambda \phi + [ (S)_{\nu, k} + R ^{(1)} + R ^{(2)} + 2 W  ] \phi  = 0 
 \eqno (4.33) $$ 
The square bracket in (4.33) is nothing but $ (-N'')_{\nu k}$.

Remark: $R \zer, R^ {(1)} R^ {(2)} $ do not involve the mutual interaction.
In contradistinction $W$ is defined through (4.24) and is model dependent 
in this sense that it crucially depends on the form of the function $f$ which 
determines the mutual interaction.

Owing to the abundance of terms depending on $\epsilon$ we distinguish 
the {\it "motional"\/}  case, where $\epsilon \not= 0$,  from the case 
{\it   "at rest"\/}  characterized by the vanishing of $\epsilon$ (or 
equivalently of $k\st$).
Notice that $ R ^{(2)} $ doesnot depend on $k \st$. It has  the 
same form in motional case and  in the case at rest [\pilk ].

\medskip
   
Let us evaluate $W$. In view of (4.24) we have first to write down the 
expression for $V''$, say (3.15). 
It follows  that 
$$ W= f(   (\widehat Z ^ \sharp )_{\nu, k} , \ 
     (P^ \sharp )^ 2 _ {\nu , k} ,  \   \nu )               \eqno (4.34)    $$ 
 In this  formula $ (P^ \sharp )^ 2 $ is given 
by (3.6) and  $\widehat Z ^ \sharp $ by (3.16).  Making the substitutions
$P \rightarrow  k$ and  $ y \sl \cdot P \sl  \rightarrow  \nu $ we obtain
$$  (\widehat Z ^ \sharp )_{\nu, k} =
  (P^ \sharp )^ 2 _ {\nu , k} \   ( \varpi z)^ 2 - (z \st \cdot k)^ 2
      \eqno (4.35)    $$
$$  ( {P^ \sharp}  ^ 2) _ {\nu, k} =   k^ 2 +
e  \    k \cdot F \cdot z    +{e^ 2 \over 4}(F \cdot z)^ 2     
  \eqno  (4.36)  $$
It is clear that $W$ does not involve the operator $z \cdot k \sl$.
Formulas  (4.35)(4.36) are to be  inserted into (4.34), then the explicit 
form of $W$ will come out.  This last expression, together with 
(4.30)(4.31)(4.32) provides the explicit form of the eigenvalue equation 
(4.33), where 
 $\lambda $  is the   eigenvalue of an 
operator  $-(N)_{\nu , k}$ acting in $L^2 (k \sl) $.

Still we meet a complication: 
just  like in the case of an isolated system, we can solve (4.13) for $k^2$ 
(condition (4.10) ensures that relation  (4.13) can be uniquely inverted)
and insert the result into  $ (N'')_{\nu , k} $. Similarly  $k\sl ^2$ can be  
replaced by 
$ \displaystyle      { k^2 \over 1- \epsilon ^2}$.
As a result eq. (4.33) bears an extra dependence on $\lambda$. In general 
it is a {\it nonconventional eigenvalue equation\/}.

Naturally a scalar product for the reduced wave function has to be explicitly 
defined.
This question will be discussed  in next section,
 in  analogy with the line we have followed in the case of an isolated system.

 According to relation (4.13), a discrete  spectrum 
for $\lambda $ would imply that the admissible values of $k^2 $ are 
restricted to a discrete sequence.

  \medskip
   {\bf 4.3. Discussion}

           Finally the system (2.21)(2.22) has been reduced to the  
three-dimensional problem of solving (4.33). This formula is nonlinear 
in the field strenght and may be applied to strong fields. 
 Let us analyze the various contributions it contains.
We distinguish motional terms, depending on $\epsilon$ or $k\st$.
In fact we can write $k\st =  \epsilon \Lambda  k\sl $ where the second rank 
tensor $\Lambda$ represent the boost from the direction of $k\sl$ to the 
direction of $k\st$ (thus $\Lambda \cdot \Lambda = \delta$).

   Loosely speaking we could say that,
in as much as the shape of  $W$ departs from the original form assumed 
by  $V  \zer$,  the mutual interaction is "modified by the  magnetic field".

\medskip
 
{\bf a) system at rest}

      The  particular case where  pseudomomentum is purely longitudinal
 enjoys a particular simplicity. If we assume for a moment that $k$
 coincides with $k  \sl$,
it is possible to find a frame where ${\bf k}$ vanishes whereas the 
electromagnetic field is purely magnetic. In other words the pseudo rest 
frame is also a lab frame.     We 
refer to this situation as the case  {\it at rest}.

In this case, $\varpi z = z \per , \  \varpi y = y \per$ 
and $(S)_{\nu , k} $  simply reduces to $y\per ^2$.
Naturally $L^2 (k\sl) $   coincides with $L^2 (k)$.
We notice that the second term in the r.h.s. of (4.35) vanishes, hence   
                $       (\widehat Z ^ \sharp)   /    
  {P^ \sharp} ^ 2 ) _{\nu, k}  $ 
reduces to  $z \per ^2 $.
Thus, for energy independent interactions ( $V\zer = g(Z/P^2, y \cdot P)$),
$W$ assumes the form $g (z \per ^2 , \nu) $.  In other words: 
 {\it   At rest, the magnetic field doesnot modify the mutual interaction when 
 it is not energy-dependent\/}.

                     All surviving terms in (4.33) 
can easily be identified as covariant generalizations of the usual terms 
present in the non-relativistic theory, except for a piece of $R^{(2)}$
which depends on the relative angular momentum  (formula (4.32)) and remains  
small for heavy bound states; its contribution  however might be significant 
for lignt bound states ($k^2 $ small) in a strong magnetic field.

    At {\it first order in the field strenght\/},
 the  relative motion admits no 
correction other than a term proportional to $\nu$ (indeed $F \cdot k $ 
vanishes). For equal masses  there is no departure from the 
motion of an isolated system.

\medskip
     {\bf b) motional case}

When $k\st$ is nonzero, we reckognize the  motional electric field contained 
in $F \cdot k$.

Energy-dependent interactions seem to be more  sensitive to the external 
field. As can be read off from (4.36),  the modification implied by (4.34) 
is nonlinear in $F$.  This point may become important in  strong fields. 

    Even if the interaction is independent of the energy, the presence 
of the motional term $z\st \cdot k$ in (4.35) entails a non-trivial 
difference between (4.34) and (2.2).

\bigskip
          {\bf 5. NORMALIZATION AND PERTURBATION THEORY.}

      {\bf 5.1. Normalization}

        In connexion with the rise of a reduced wave function $\phi$ 
in (2.23),  we are led to     consider in general  any function 
(or possibly a distribution)  of  $z$     of the form
$ \phi (k,z) $ which depends on $z$ only through the combination $ \varpi z$
(assuming $k$ timelike), regarding  the four-vector $k$ as a given parameter.
The most  straightforward normalization of $\phi$  is given by a  three-fold 
integral using the volume element 
   $    \ d^ 3  ( \varpi z)  $,   
with the convention that 
$  \varpi z   =   {\bf z} $ in any frame adapted to $k\sl$.
In other words, for each $k$, we separate $k \sl $ from $k$ and consider  the 
Hilbert space
  $ \displaystyle    L^ 2 (k\sl )  =  L^ 2 ({\bf R} ^ 3 , d^ 3 (\varpi z))$.

 For a function of the above type, but not necessarily solution to (4.33),
we have, with an obvious notation
$$ <\phi  , \phi  >_k   =   
     \int  \phi ^ *  \   \phi   \ d^ 3  ( \varpi z)  \eqno (5.1) $$
Here  $\phi $ may additionally depend on $\nu$ and $k \st$ 
as parameters.   The label $k$ refers to the {\it vector\/} $k$, but 
actually  $L^2 (k\sl ) $ depends on $k^\alpha$ 
only through  the direction of $k \sl $.

     It is noteworthy that the  normalization   (5.1) of reduced wave 
functions is  not only dictated by simplicity,
 but also  consistent  with the  {\it off-shell  normalization\/}
 of the full wave functions.

The proof of this point will be given elsewhere; its precise meaning is as 
follows:

Let us consider any function 
$ \Psi ''(Q, z) $ irrespective of its being on the mass shell or not.

                   Assuming that 
$ \phi (z^\alpha , k^\beta , \nu) $ 
is related to $\Psi '' $ through (2.23), it is possible to check that 
$$ \int  {\Psi '' } ^*  \Psi ''   \    d^4 Q d^4 z =
\int  <\phi ,  \phi >_k    d^4 k  d \nu                 $$
provided that the Fourier transform of $\Psi '' $ with respect to $Q$,
say  $\Xi (k,z)$,  is a  retarded function of $k$ 
(it  vanishes outside the region limited by the positive sheet of the 
light cone).

     Of course this nice property [\dixm ]
doesnot prevent the complications associated with energy dependence.

Even in the simple case where the mutual interaction $V \zer$ (present in the 
isolated system) doesnot depend on $P^2$, we are bound to realize that 
the  spectral problem {\it is  not\/} a standard one, 
due to the presence of the magnetic field.  This can be seen as follows.
First we notice that the occurence of $ ( \widehat Z ^\sharp)_{\nu , k}$
in $W$ brings out a dependence on  $k^2, k\sl ^2$.
Second  we observe         an unescapable  dependence  on $k^2$ 
and $k\sl ^2$ in formulas  (4.28)(4.31)(4.32).
Clearly, we cannot expect a conventional   spectral problem.
This peculiarity  arises because going from $\Psi$ to $\Psi ''$, we have 
abandoned the customary representation; transformation formulas (2.4)(2.19) 
are responsible for the nonlinear dependence on $\lambda$.

\bigskip
 
  {\bf 5.2 Outline of a perturbative approach.}

In order to  solve  the reduced wave equation, perturbation theory is by no 
means straightforward.
  The abundant mathematical literature devoted to nonlinear eigenvalue 
equations [\kel ] pays very little attention to spectrum perturbation.
So the general case seems to be hopeless in the present state of the art 
[\algo ].

{\it Henceforth we limit ourselves to the simple case where $V \zer $ 
is not energy dependent\/}.
Still we must cope with a nonconventional eigenvalue equation when  magnetic 
field is present.
  But we can make a couple of  important remarks.

First we observe that  all terms involving energy dependence 
vanish when {\it both\/} the external field and the motional parameter vanish. 

    Second, we notice that $\epsilon$ and $k^2$ are two independent 
parameters contributing to the determination of the vector $k ^\alpha$.
Inspection of (4.13) shows that $\epsilon$ and $\lambda$ are mutually 
independent.

These remarks open the possibility to {\it treat simultaneously  
 electromagnetic contributions and motional terms as a perturbation}.
This procedure  requires that the motional parameter 
is not too large.

In this approach  the unperturbed equation  is free of nonlinear dependence on 
$\lambda$.
It corresponds to "static states" of 
the isolated system, {\it i.e.\/} states with external coupling removed and 
additionally at rest  with respect to the special lab frame; they are 
characterized by $k\st = 0, \   F=0 $ (see Appendix). 
Only the perturbation, which combines motional terms and field contributions, 
is affected by energy dependence.
It is fortunate that Rizov,  Sazdjian and Todorov [\rst ] have developed a 
perturbative scheme (tractable by physicists) for this situation.

\medskip
 
Another departure from conventional perturbation theory lies in the 
fact that  the perturbation is nonlinear in $\epsilon$ and 
in the field strenght.

Indeed, insofar as $f$ in (2.6) can be developed as an analytic function of its 
arguments,  a look at (4.35)(4.36) indicates that 
 even  if  $f$ is linear in $Z$ (harmonic potential), $V''$ and $W$ are
 at least quadratic in the external field.

     Fortunately, the formalism  set up in [\rst ] is general enough to 
accomodate second (and higher) order perturbations.

\medskip
 
The first step is  a {\it weak-field-slow-motion approximation\/}.
The field strenght and $\epsilon$ are taken into account only at first order.
The perturbation scheme of ref. [\rst ] gets simplified; it turns out that 
the correction to the eigenvalue is given by the usual formulas. In 
particular if the unperturbed eigenvalue is nondegenerate, the correction 
is still given by the expectation value of the perturbation term in the 
eigenstate of the unperturbed problem.
We can organize calculations by setting 
$F / \mu    = \epsilon G $ 
(or alternatively  $ F / k^2 = \epsilon G $) where the tensor 
$G$ is dimensionless. It follows that $F \cdot k$ and  
the motional electric field are  $ O(\epsilon ^2) $.
The first order treatment entails enormous simplifications.
For instance the obligation to distinguish  $ (z \per )^2$ from  
$ (\varpi z )^2 $  is in general a serious computational complication,
but these quantities coincide at first order in  $\epsilon$.
A glance at (4.35) shows that 
$W = f ( z\per ^2 , \nu) + O (\epsilon ^2) $. 
At first order, the only 
departure from the unperturbed equation consists in the $\nu$-terms and 
vanishes  for equal masses. To summarize:

{\it If the   neutral  system as a whole undergoes a slow motion in a 
weak field, the relative motion is affected by the magnetic field only 
by a contribution which vanishes for equal masses\/}.

    Therefore it is  necessary to go beyond first order in search for a 
nonzero correction.

{\it Example:}

Suppose that $V \zer = g(\widetilde z ^2 )$. This form of mutual interaction 
includes the  relativistic harmonic oscillator, 
say  $V \zer = \gamma  Z / P^2 $, where $\gamma $ is a coupling constant. 
The only difference from the unperturbed equation comes from 
  $R ^{(1)} $, which is given by (4.31).

\medskip
            Another situation of great simplicity consists in a system of 
equal masses at rest ($\epsilon $ and $\nu$ are strictly zero).
The perturbation reduces to $R^{(2)}$, hence is linear in $|F|^2 $. Corrections 
to the levels are easy to calculate, but  significant only in strong fields.

\vfill \eject

{\bf 6.  CONCLUSION.}

The coupled Klein-Gordon equations describing a neutral system 
 have been  reduced to a three-dimensional eigenvalue equation
involving truly motional effects and recoil effects in a   covariant fashion.
Moreover the particular symmetry associated with a constant magnetic field
{\it in space-time\/}  is manifestly respected.

Explicit formulas are written in a representation which permits to 
easily 
satisfy compatibility and also to eliminate relative time and the $Q$-
variables.  The surviving degrees of freedom are the same as in the 
nonrelativistic theory.

In the reduction procedure it was essential to consider eigenstates of 
pseudomomentum.
The square of this vector plays the role of an effective squared 
mass which can be, in principle, evaluated by solving the reduced equation.
Indeed  (in the sector we have considered, and especially for equal masses)
  the eigenvalue $\lambda$ is in one-to-one correspondance with  $k^2$. 
Bound states are characterized by $\lambda $ in the discrete spectrum.

    In the most natural way, our approach  generalizes to the magnetic 
case the usual treatment of isolated  two-body systems according to 
"predictive mechnics" 
and "constraints theory" [\rep ][\coupl ][\sazdj ][\sazjmp ].
 It provides a clean theoretical basis for the study of neutral bound states 
in the relativistic regime.

\medskip
 
The equations of motion  are nonlinear in the field strenght and offer
 a starting point for investigation of strong field effects.
In principle, they encompass all kinematic possibilities of the system as a 
whole and   permit a description of ultra-relativistic situations.
The Ansatz which allows for a three-dimensional reduction in the covariant 
framework  automatically generates various terms in the wave equation.
From a practical point of view, 
some of them play the role of a modification of the mutual interaction, 
 caused by the magnetic field.
This effect might have dramatical consequences in a system where the mutual 
interaction is energy-depending (for instance Coulombian interaction along 
the lines of the quasi-potential approach).
 But in this case we cannot go beyond  
qualitative estimations, since the  perturbative scheme borrowed from ref. 
[\rst ] is limited to the 
systems where the interaction is not energy dependent.   This point may be a 
 motivation for   eventually undertaking a more general perturbation theory
 applicable also to  eigenvalue equations involving an energy-dependent 
potential.

\medskip
   When $V \zer $ does not depend on $P^2$,
spectrum calculations  can be handled in a perturbative 
approach where the motional parameter as well as the field strenght
must necessarily be small.
For the moment this method can be used for simple and idealized systems 
bounded by a phenomenological potential.
Application to the relativistic "naive quark" model of hadron
seems to indicate that the spectrum  enjoys some kind of stability with 
respect to magnetic perturbation. 
Of course the polarizability of this model can be studied in our formalism.

\medskip
  In the hope of obtaining  a nonvanishing  
correction due to the presence of external field, we are obliged to go beyond 
first order. In a future work we plan to investigate with more details the 
case  $\epsilon=\nu=0$ sketched in last section.

    More work is needed  in search for  ultra relativistic effects 
( $\epsilon \simeq 1 $). 
A less ambitious program might be the computation of second order 
 motional effects in the framework of the perturbative scheme considered here.
But the complication of the calculations may be prohibitive.

Naturally an extension to particles with 
 spin is desirable.

Let us finally mention that the contact with usual methods of quantum field 
theory could be improved  by  considering a  Bethe-Salpeter  equation  taking
 the magnetic field into account from the start.
 This would 
    mean to resume the  work of Bijtebier and Broeckaert [\brok] in a way which 
    respects the particular symmetry of constant magnetic field, $ i.e.$ 
    treating all the possible lab frames on the same footing.

\bigskip
 
\centerline {====================================}

    \noi
    \centerline  {\bf APPENDIX}

\noi
{\bf The unperturbed equation}

\noi   The vanishing of $k\st $ (equivalently of $\epsilon$) makes $L^2 (k\sl) $
to coincide with  $ L^2 (k) $ and $\varpi z$ with $z_\perp $.
As also  $F$ is zero, inspection of (4.34)(4.35)(4.36) shows that $W$ reduces 
to   $ \displaystyle   
   f(( Z)_{\nu , k} , k^2 , \nu )  =
              f ( z^2 k^2 -  (z \cdot k) ^2 , k^2 , \nu )       $
Finally we see that, for $\epsilon = F = 0$,  (4.33) reduces to
  $$ \lambda \phi +  z _ \perp ^2  \phi  +
   f ( z^2 k^2 -  (z \cdot k) ^2 , k^2 , \nu ) \  \phi     = 0       $$        
It is exactly the equation one would obtain for the isolated system after 
reduction, in the original representation. 

\vfill  \eject  

\noi
   \item {1}     Ph.DROZ-VINCENT, Nuovo Cimento A, {\bf 105}, 1103 (1992).
      We are speaking of frames defined up to an arbitrary space 
   rotation. Such a frame essentially corresponds to a timelike  direction.

\noi
   \item {2}  J.E.AVRON, I.W. HERBST, B. SIMON, Ann. Phys. (N.Y.) {\bf 114},
 431 (1978)

\noi
   \item {3}  M.H.JOHNSON, B.A.LIPPMAN, Phys.Rev. {\bf 76}, 828 (1949)

  \noi
 \item {4}   H. GROTCH and R.A. HEGSTROM,  Phys.Rev. A {\bf 4}, (1971) 59.

\item {\dzy }          L.P.GOR'KOV  and  I.E.DZYALOSHINSKII, 
Soviet Phys. JETP {\bf 26}, 449 (1968).

\noindent
   \item {\rep}   Ph.DROZ-VINCENT,  Reports in Math. Phys. {\bf 8}, (1975) 79.
Phys.Rev.D {\bf 19}, 702 (1979)

 \noindent
   \item {\coupl}   Coupled wave equations have been considered also by
L.BEL,  in Differential Geometry and Relativity, M.Cahen and M.Flato editors,
Reidel Dordrecht (1976) 197, Phys. Rev. D {\bf 28} (1983) 1308.
  H.LEUTWYLER and J.STERN, Ann.of Phys.(N-Y){\bf 112} (1978) 94,
  Phys.Lett.B  {\bf 73} (1978) 75. 
  H.CRATER and P.VAN ALSTINE  Phys.Lett.B {\bf 100} (1981) 166.

\noi
\item {\sazjmp}            Realistic cases are more complicated. 
               See for instance TODOROV in   ref.6,
            or H.SAZDJIAN, J.Math.Phys. {\bf 28}, 2618 (1987). 

 \noindent
   \item {\bij89  }   J.BIJTEBIER,  Nuovo.Cim.A {\bf 102}, 1285 (1989)

\noi
\item {\dv95 }  Ph. DROZ-VINCENT, Phys.Rev. A {\bf 52}, 1837 (1995)

\noi
\item {\purmagn} We say that a field is pure electric (resp. pure magnetic) 
when there exist frames where the field appears to be so; of course these two 
situations exclude one another.

 \noi
\item {\rst}   V.A.RIZOV, H.SAZDJIAN and I.T.TODOROV, Ann. of Phys. {\bf 165}, 
59-97 (1985). 

\noi
 \item {\rigo}    For a more rigorous exposition we should start with 
$ \Psi ' $ from the outset. See Ph.DROZ-VINCENT,
   Few-Body Systems, {\bf 14}, 97-115 (1993)

\noi
   \item {\L.L }     With this notation be cautioned  that 
           $L \cdot L  \not=  L^2 $.          For all vectors
        $\xi , \eta $ we write 
   $\xi \cdot    F  \cdot \eta $ for   
   $\xi ^\alpha     F _{\alpha \beta}   \eta ^\beta  $  

\noi
\item {\anal }  Ph.DROZ-VINCENT, Ann. Inst. H. Poincar\'e, {\bf 27}, 407-424
 (1977)

  \noi
 \item {\sazdj} H.SAZDJIAN, Phys.Rev.D {\bf 33}, 3401 (1986).

\noi
\item {\klambda } 
It may be tempting to directly consider $k^2$ as the eigenvalue in the 
reduced equation.
This interpretation would be  unfortunate because,
 even in the simple case where $V$ is not energy dependent,
  for unequal masses, $k^2 $ arises in a nonlinear manner (for equal masses,
 the point becomes academic).
Moreover $\lambda$ arises naturally as eigenvalue of $-N$.
As well as in nonrelativistic mechanics, it is convenient to associate the
spectrum with relative motion, and $N$ is precisely related to relative energy.

\noi
\item {\kel }   M.V.KELDYSH, Dokl. Akad. Nauk SSSR {\bf 77}, 11-14 (1951).
Uspekhi Mat. Nauk {\bf 26}, 15-41 (1971)=Russian Math. Survey, {\bf 26}, 14-44 
(1971).
G.V.RADZIEVSKII, Uspekhi  Mat. Nauk {\bf 37}, 81-145 (1982)
                =Russian Math. Surveys {\bf 37}, 91-164 (1982).

\noi
\item {\wein}  H.SNYDER and J.WEINBERG, Phys. Rev. {\bf 57}, 307-314 (1940).
L.I.SCHIFF,   H.SNYDER and J.WEINBERG, Phys. Rev. {\bf 57}, 315-318 (1940).

\noi
\item {\pilk }  The rest case, in strong fields, has been earlier
  considered by   D.F.KOLLER, M.MALVETTI and H.PILKUHN, 
 Phys.Lett. {\bf 132} A, 259 (1988).    
  Motional effects have been considered in the noncovariant litterature, see
M.MALVETTI and H.PILKUHN, Phys. Reports {\bf 248}, 1-60 (1994), and 
references therein. 

\noindent                                                          
\item {\dixm}     The family of Hilbert spaces  $L^2 (k \sl)$
         is supposed to be  labelled by the  "continuous indices" $k^\alpha$
         and $\nu $   but it  is a "constant field" with  respect to $\nu $
      and to  the transverse parameters forming the components of  $k \st $).  
 For this terminology see J.Dixmier, "Les alg\`ebres 
d'op\'erateurs dans l'espace Hilbertien", Gauthier-Villars, Paris (1969) 
chap.II. 

\noi 
 \item {\algo} At first order however, 
                    a "heuristic algorithm" can be carried 
out, setting  $  \lambda  =       \lambda \zero   +   \lambda _1  $  and 
$  \phi  =  \phi \zer    +  \phi _1 $.
Inserting into the wave equations and neglecticting the squares of $ \lambda 
_1, \phi _1 $ automatically yields a correction $\lambda _1$, provided we 
impose 
$ \int \phi _1 ^*  \phi \zero \  d^3( \varpi z)= 0  $. 
Needless to say, this  procedure might be unfounded if the spectral theory of 
the unperturbed problem is not previously under control.

\noindent
   \item {\brok }  J.BIJTEBIER and J.BROEKAERT, Nuovo Cimento Soc.Ital.Phys. A,
 {\bf 105}, 351  (1992).

\end